\documentstyle[12pt]{ioplppt}
\begin{document}
\title{Generalised equilibrium of cosmological fluids in second-order thermodynamics}

\author{Winfried Zimdahl\dag\ddag, J\'{e}r\^{o}me Gariel\dag and G\'{e}rard Le Denmat\dag}

\address
{
\dag\ Laboratoire de Gravitation et Cosmologie Relativistes, 
Universit\'e Pierre et Marie Curie- CNRS/URA 769, 
Tour 22, 4\`eme \'etage, Bo\^{i}te 142 
4, 
place 
Jussieu\\ 
75252 Paris Cedex 05, 
France
}

\address{\ddag\ Fakult\"at f\"ur Physik, Universit\"at Konstanz, 
PF 5560 M678
D-78457  Konstanz, Germany\footnote{present address}
}


\begin{abstract}
Combining the second-order entropy flow vector of the causal Israel-Stewart theory with the conformal Killing-vector property of $u _{i}/T$, 
where $u _{i}$ is the four-velocity of the medium and $T$ its equilibrium temperature, 
we investigate generalized equilibrium states for cosmological fluids with nonconserved particle number. 
We calculate the corresponding equilibrium particle production rate 
and show that this quantity is reduced compared with the results of 
the previously studied first-order theory. 
Generalized equilibrium for massive particles turns out to be compatible 
with a dependence 
$\rho \propto a ^{-2}$ of the fluid energy density $\rho $ on the scale factor $a$ of the Robertson-Walker metric and may be regarded as a realization of 
so-called K-matter. 
\end{abstract}

PACS numbers: 98.80.Hw, 95.30.Tg, 04.40.Nr, 05.70.Ln

\section{Introduction}
Among the theoretical efforts to understand the physics of the early universe 
the investigation of cosmological nonequilibrium processes is 
of considerable interest (see, e.g., \cite{KoTu,Boe}). 
A characteristic feature of  many nonequilibrium phenomena is the 
nonconservation of particle numbers in interaction processes at high energies within the cosmic substratum or due to quantum particle production. 
Generally, cosmological particle creation has been extensively studied in  quantum field theory in curved spacetime \cite{BiDa,Ford}. 
Progress in nonequilibrium quantum field theory and its cosmological applications in the context of fluctuation-dissipation theorems was obtained in 
\cite{HuSi,CaVe} and references therein. 
As was realized by Zel'dovich \cite{Zel}, Murphy \cite{Mur} and 
Hu \cite{Hu}, a quantum particle production rate may phenomenologically be equivalent to an effective viscous pressure. 
This correspondence has initiated a series of papers discussing a fluid phenomenological approach to cosmological particle production 
\cite{Prig,Barr,Calv,LiGer,ZP1,ZP2,ZP3,GaLeDe,Lima,WZ,GunzMaNe,JeGer}. 
Special attention has been devoted to ``isentropic'' 
(or ``adiabatic'') particle production, characterized by a given, constant equilibrium entropy per particle. In such a case the entropy production associated with the creation of particles 
is entirely due to the enlargement of the phase space of the fluid particle system and not to fluid dissipative processes which would also 
enlarge the entropy per particle. 
Particle creation is an intrinsically nonequilibrium process. 
While a full out-of-equilibrium treatment is beyond the present theoretical possibilities, we expect the concept of ``generalized'' equilibrium,   previously introduced by one of the authors \cite{ZGeq}, to be a step in this direction. 
``Generalized'' equilibrium states share essential features with ``conventional'' equilibrium states. They are, on the other hand, compatible with a nonvanishing entropy production. 
More specifically, 
this kind of equilibrium relies on the existence of a conformal, timelike Killing vector and is compatible with nonzero ``isentropic'' 
fluid particle production. 
Despite of the change of the particle number it is characterized by 
a J\"uttner distribution function for collisional equilibrium on the level of kinetic theory \cite{TZP,ZTP,ZiBa}. 
For massless particles the equilibrium conditions require the production rate to vanish and the ``generalized'' equilibrium coincides with the familiar ``global'' equilibrium of standard relativistic thermodynamics 
in this limit. 
Particles with nonzero mass for which a ``global'' equilibrium in the standard sense cannot generally exist 
in an expanding universe, may be in ``generalized'' equilibrium provided there is net 
production of these particles. 
The corresponding production rate depends on the fluid equations of state and is fixed by the equilibrium properties of the fluid. 
It is highest for nonrelativistic matter in which case it amounts to half the expansion rate. 
In case of spatial homogeneity 
such kind of equilibrium requires accelerated expansion, i.e., it can 
only be realized in a power-law inflationary universe. 
This result represents a surprising link between equilibrium properties of the cosmological fluid and the expansion behaviour of the universe. 
While the ``generalized'' equilibrium concept is closely related to the properties of a timelike conformal Killing vector it is also based on a specific expression for the entropy flow vector. 
The latter was assumed to be of the Eckart type in \cite{ZGeq} taking into account first-order deviations from conventional equilibrium. 
It is well known \cite{Mue,I,IS,PJC-V,HL} that the first-order Eckart theories suffer from serious drawbacks concerning causality and stability. 
If one includes higher-order deviations from equilibrium as well, these  problems disappear. 
By now, it is generally agreed that any analysis of out-of-equilibrium 
phenomena in relativity should be based on the theories of M\"{u}ller \cite{Mue}, Israel \cite{I}, Israel and Stewart \cite{IS}, or Pav\'{o}n, Jou  and Casas-V\'{a}zquez \cite{PJC-V}, 
including at least second-order deviations fom (conventional) equilibrium. 
These theories are ``causal'' in the sense that dissipative signals propagate only at subluminal speeds (see \cite{APR} for a recent review). 

It is the aim of this paper to extend the ``generalized'' equilibrium concept to the second-order theory 
and to investigate the cosmological dynamics for the case that the matter content of the universe is in such kind of equilibrium.  
We will restrict ourselves to (effective) bulk pressures throughout this paper. 
It should be emphasised that the restriction to only one entropy producing phenomenon is more severe in a second-order theory than it is in Eckart's theory. 
In general, the causal evolution equations for the viscous pressures and the heat flux are coupled. 
Consequently, our considerations apply to cases where this coupling is not essential. 
In spatially homogeneous spacetimes, the main case of interest here, there is no coupling at all and entropy production is only possible due to nonvanishing bulk pressures. 
It will turn out that the particle production rate required by the generalized equilibrium conditions of the second-order theory 
is lower than the corresponding rate in the first-order theory. 
This strongly reduces the possibility of accelerated expansion (power-law inflation) in a spatially homogeneous universe as a consequence of the back reaction of equilibrium particle production on the cosmological dynamics. 
The generalized equilibrium conditions of the second-order theory may be compatible, however, with a ``coasting'' cosmology \cite{Kolb}, i.e., a universe expanding with constant velocity. 

The paper is organized as follows. 
In section 2 we introduce the concept of generalized equilibrium 
in the context of second-order theories of fluid dynamics. 
For later comparison section 3 briefly recalls the corresponding properties 
of the first-order theory. 
Section 4.1 discusses the relation between the generalized equilibrium conditions and the causal evolution equation for the viscous pressure in the Israel-Stewart theory.  
Section 4.2 demonstrates that the second-order theory admits equilibrium states with nonvanishing viscous pressure even for conserved particle number 
in case $u _{i}/T$ is a conformal Killing vector. 
In section 4.3 our main results concerning the 
generalized equilibrium production rate for massive particles 
in the expanding universe are obtained 
and discussed.  
Section 5 presents the conclusions of the paper.  
Units have been chosen so that $c = k_{B} = 1$. 

\section{The concept of ``generalised equilibrium''}
In order to introduce the generalised equilibrium concept, we first recall the  standard equilibrium properties of relativistic gaseous fluids. 
A relativistic gas of particles with elastic binary interactions, described by Boltzmann's collision integal, is at (local) equilibrium if the corresponding one-particle distribution function is of the J\"uttner type. The collision integral vanishes under such circumstances. 
If, moreover, this equilibrium distribution is supposed to obey Boltzmann's 
equation, the gas is said to be at ``global'' equilibrium. 
For particles with nonvanishing rest mass this requires the existence of a timelike Killing vector, i.e., stationarity. 
Such kind of equilibrium cannot exist in an expanding universe. 
For massless particles global equilibrium is realised if there exists a conformal timelike Killing vector. This is of cosmological relevance since it is compatible with a nonvanishing expansion rate. 
The concept of ``generalised'' equilibrium is an extension of the global equilibrium concept insofar as it takes into account additional interactions within the gaseous fluid which, in general, are not elastic and result in a change of the particle number. 
However, it retains essential features of the global equilibrium, namely both the structure of the equilibrium distribution function which makes Boltzmann's collision integral vanish (at least within the first-order theory) 
and the conformal Killing vector condition for $u _{i}/T$. 
As a consequence, such kind of equilibrium becomes possible also for particles with nonvanishing rest mass, provided that the particle number increases at a certain rate. 
In the present paper we shall investigate the implications of generalised equilibrium for the second-order theory on a phenomenological level. 
Under these premises we start our considerations with the energy-momentum tensor of a fluid with bulk stresses, 
\begin{equation}
T ^{ik} = \rho u ^{i}u ^{k} + \left(p + \pi  \right)h ^{ik} \ .
\label{1}
\end{equation}
Here, $\rho $ is the energy density of a fiducial equilibrium state, $p$ is the corresponding equilibrium pressure, $u ^{i}$ is the fluid four-velocity in the Eckart frame and $h ^{ik} = g ^{ik} + u ^{i}u ^{k}$ is the spatial projection tensor. 
The quantity $\pi $ denotes that part of the scalar pressure which is connected with entropy production. 
The local energy-momentum conservation $T ^{ik}_{\ ;k} = 0$ implies the balances 
\begin{equation}
\dot{\rho } + \Theta \left(\rho + p + \pi  \right) = 0
\label{2}
\end{equation}
and 
\begin{equation}
\left(\rho + p + \pi  \right)\dot{u}_{m} 
+ \nabla  _{m}\left(p + \pi  \right) = 0\ ,
\label{3}
\end{equation}
where $\dot{\rho } \equiv  \rho _{,i}u ^{i}$ etc. and 
$\nabla  _{m}p \equiv   h ^{n}_{m}p _{,n}$. 
Within the Eckart frame the particle number flow vector $N ^{i}$ is given by 
\begin{equation}
N ^{i} = n u ^{i}\ ,
\label{4}
\end{equation}
where $n$ is the particle number density. 
We will not assume that the fluid particle number is conserved. Instead, we admit a source term $\Gamma $ in the corresponding balance equation which describes the rate of change of the fluid particles:
\begin{equation}
N ^{i}_{;i} = n \Gamma \ .
\label{5}
\end{equation}
Inserting here the expression (\ref{4}) yields
\begin{equation}
\dot{n} + \Theta n = n \Gamma \ ,
\label{6}
\end{equation}
where $\Theta \equiv  u ^{i}_{;i}$ is the fluid expansion.  
The quantity $\Gamma $ describes the net change of the particle number either due to internal reactions within the medium \cite{ZMNRAS} or due to particle production in strongly varying gravitational fields 
\cite{BiDa,Zel,Hu}. 
Usually, a quantity such as $\Gamma $ 
is an input quantity in a phenomenological description and has to be calculated from the underlying microphysics. 
Here, it will largely be fixed by the generalized equilibrium conditions. 

Limiting ourselves to second-order deviations from equilibrium, the entropy flow vector $S ^{a}$ takes the form \cite{IS,HiSal}
\begin{equation}
S ^{a} = ns u ^{a} - \frac{\pi ^{2}}{2}\frac{\tau }{\zeta }
\frac{u ^{a}}{T}\ ,
\label{7}
\end{equation}
where $s$ is the entropy per particle and $T$ is the temperature. 
The quantity $\tau $ plays the role of a relaxation time while $\zeta $ 
is identified with the coefficient of bulk viscosity in conventional second-order theories. 
Covariant differentiation of the expression (\ref{7}) yields
\begin{equation}
S ^{a}_{;a} = s N ^{a}_{;a} + n \dot{s} 
- \frac{1}{T}\left(\frac{\pi ^{2}}{2}
\frac{\tau }{\zeta } \right)^{\displaystyle \cdot} 
- \frac{\pi ^{2}}{4}\frac{\tau }{\zeta }g ^{ab}
\left[\left(\frac{u _{a}}{T} \right)_{;b} + 
\left(\frac{u _{b}}{T} \right)_{;a}\right]\ .
\label{8}
\end{equation}
From the Gibbs equation (see, e.g., \cite{Groot})
\begin{equation}
T \mbox{d}s = \mbox{d} \frac{\rho }{n} - p \mbox{d} \frac{1}{n}
\label{9}
\end{equation}
one obtains 
\begin{equation}
n T \dot{s} = - \Theta \pi - \left(\rho + p \right)\Gamma  \ .
\label{10}
\end{equation}
With the help of Eq. (\ref{10}) the equation for the entropy 
production density (\ref{8}) may be written as
\begin{equation}
S ^{a}_{;a} - n \Gamma s = - \frac{\Theta \pi }{T} 
- \frac{\rho + p}{T}\Gamma 
- \frac{1}{T}\left(\frac{\pi ^{2}}{2}
\frac{\tau }{\zeta } \right)^{\displaystyle \cdot} 
- \frac{\pi ^{2}}{4}\frac{\tau }{\zeta }g ^{ab}
\left[\left(\frac{u _{a}}{T} \right)_{;b} + 
\left(\frac{u _{b}}{T} \right)_{;a}\right]\ .
\label{11}
\end{equation}
On a phenomenological level, ``generalized equilibrium'' is defined as the vanishing of 
$S ^{a}_{;a} - n \Gamma s$ together with the conformal Killing-vector property of $u _{i}/T$ \cite{ZGeq}: 
\begin{equation}
S ^{a}_{;a} = n \Gamma s \ , 
\mbox{\ \ \ } \Gamma \geq 0 \ ,
\mbox{\ \ \ }
\left(\frac{u _{i}}{T} \right)_{;k} 
+ \left(\frac{u _{k}}{T} \right)_{;i} 
= 2 \phi g _{ik}\ .
\label{12}
\end{equation}
The term $n \Gamma s$ on the right-hand side of the first relation 
in Eq. (\ref{12}) 
takes into account the increase of the entropy of the system due to the enlargement of the phase space. 
Simply due to its very existence each of the created particles contributes an equilibrium amount $s$ to the entropy of the fluid system. 
If there is production of entropy only due to the enlargement of the phase space where each particle contributes an equilibrium amount $s$ but not due to dissipative processes, the system will be considered to be in ``generalized'' equilibrium. 
Within the first-order theory 
the microscopic constituents of such a fluid are governed by a distribution function for collisional equilibrium \cite{TZP,ZTP,ZiBa}. 
For $\Gamma = 0$ the conditions (\ref{12}) reduce to the standard 
equilibrium conditions of relativistic thermodynamics. 
The conformal Killing-vector 
(CKV) property of $u _{i}/T$ implies the relations 
\begin{equation}
\phi = \frac{1}{3}\frac{\Theta }{T}\ , 
\mbox{\ \ \ \ \ \ \ \ \ \ \ \ }
\frac{\nabla  _{a}T}{T} + \dot{u}_{a} = 0 \ ,
\label{13}
\end{equation}
and
\begin{equation}
\frac{\dot{T}}{T} = - \frac{\Theta }{3}\ .
\label{14}
\end{equation}
We point out that the second relation (\ref{13}) implies the vanishing of the heat flux in Eckart's theory, equivalent to a vanishing of the corresponding first-order contribution to the entropy production within the Israel-Stewart theory. In homogeneous spacetimes this relation is trivially fulfilled. 

We will generally assume that the equilibrium variables $p$ and $\rho $ obey equations of state in the form 
\begin{equation}
p = p \left(n,T \right)\ ,\ \ \ \ \ \ \ \ \ 
\rho = \rho \left(n,T \right)\ .
\label{15}
\end{equation}
Differentiating the latter relation and using 
the balances (\ref{2}) and  
(\ref{6}) as well as Eq. (\ref{10}) we obtain \cite{Calv,LiGer,ZP1}
\begin{equation}
\frac{\dot{T}}{T} = - \left(\Theta - \Gamma  \right) 
\frac{\partial p}{\partial \rho } 
+ \frac{n \dot{s}}{\partial \rho / \partial T}\ .
\label{16}
\end{equation}
Here we have used the general relation 
\begin{equation}
\frac{\partial{\rho }}{\partial{n}} = \frac{\rho + p}{n} 
- \frac{T}{n}\frac{\partial{p}}{\partial{T}}\ ,
\label{17}
\end{equation}
which follows from the fact that the entropy is a state function and the abbreviations 
\[
\frac{\partial{p}}{\partial{\rho }} \equiv  
\frac{\left(\partial p/ \partial T \right)_{n}}
{\left(\partial \rho / \partial T \right)_{n}}\ ,
\ \ \ \ \ \ \ \ \ 
\frac{\partial{\rho }}{\partial{T}} \equiv  
\left(\frac{\partial{\rho }}{\partial{T}} \right)_{n} \ ,
\ \ \ \ 
{\rm etc.}
\]
 
The temperature laws (\ref{14}) and (\ref{16}) have to be consistent. 
This requirement fixes the production rate. 
For later comparison we first review the results of the first-order theory 
\cite{ZGeq}, equivalent to $\tau = 0$ in (\ref{11}). 

\section{The first-order theory}
In the first-order theory the first ``generalized'' equilibrium condition (\ref{12}) reduces to (cf (\ref{11}) with $\tau = 0$)
\begin{equation}
\Theta \pi = - \left(\rho + p \right)\Gamma \ , 
\mbox{\ \ \ \ \ \ \ \ \ \ \ \ \ \ \ } 
\left(\tau = 0 \right)\ ,
\label{18}
\end{equation}
equivalent to $\dot{s} = 0$ (cf (\ref{10})). 
For $\Gamma = 0$ this requirement reduces to the condition $\pi = 0$, 
a case dealt with by Bedran and Calv\~{a}o \cite{BedCalv}. 

Comparing now the temperature laws (\ref{14}) and (\ref{16}) yields 
\begin{equation}
\Gamma = \left(1 - \frac{1}{3}\frac{\partial \rho }
{\partial p} \right)\Theta \ , 
\mbox{\ \ \ \ \ \ \ \ \ \ \ \ \ \ \ }
\left(\tau = 0 \right)\ .
\label{19}
\end{equation}
The CKV property of $u _{i}/T$ fixes the creation rate. 
An equilibrium according to $S ^{a}_{;a} - ns \Gamma = 0$ is only possible 
if fluid particles are created at a specific rate given by Eq. (\ref{19}). 
The production rate $\Gamma $ vanishes for $p = \rho /3$, i.e. for 
radiation. 
This means recovering the well-known fact that a fluid obeying an  
equation of state $p = \rho /3$ may be in ``global equilibrium'' if the spacetime admits a 
timelike CKV. 
For any other equation of state $0 < p < \rho /3$ 
there exists a specific production 
rate (\ref{19}) which is necessary to fulfill the equilibrium conditions. 
The production of particles at a rate different from the rate  
(\ref{19}) disturbs the equilibrium. 
Except for $p = \rho /3$ there is no equilibrium of 
this kind for $\Gamma = 0$. 
In other words, a fluid with an equation of state in the range 
$0 < p < \rho /3$ may only be in equilibrium if fluid particles are produced 
at a specific rate depending on the equation of state. 

Inserting the relation (\ref{19}) into the condition 
(\ref{18}) fixes $\pi $:
\begin{equation}
\pi = - \left(\rho + p \right)
\left(1 - \frac{1}{3}\frac{\partial \rho }
{\partial p} \right) \ ,
\mbox{\ \ \ \ }
\left(\tau = 0 \right)\ .
\label{20}
\end{equation}
The equilibrium conditions naturally imply the existence of 
an effective bulk pressure, i.e., the latter has not to be postulated separately. 
It is obvious that $\pi $ as well as $\Gamma $ vanish for $p = \rho /3$.  
For nonrelativistic matter, however, characterized by equations of state 
$p = nT$ and $\rho = nm + \frac{3}{2}nT$ with $m \gg T$, one obtains 
\begin{equation}
\Gamma = \frac{1}{2}\Theta \ , \mbox{\ \ \ }€
\pi = - \frac{1}{2}\left(\rho + p \right) 
\approx - \frac{\rho }{2}
\mbox{\ \ \ \ \ \ \ \ \ \ \ \ \ \ \ }
\left(m \gg T \ ,\  
\tau = 0 \right)\ .
\label{21}
\end{equation}
The rate $\Gamma $ is half the expansion rate and $|\pi|$ is a remarkable 
fraction of the energy density. 
With equations of state of the general form (\ref{15}) the particle number density $n$ and the temperature $T$ are the primary thermodynamic variables 
of our system. 
While the equilibrium behaviour of the temperature is given by 
Eq. (\ref{14}) independently of the equation of state, the particle number density changes according to 
\begin{equation}
\frac{\dot{n}}{n} = - \frac{1}{3} \frac{\partial{\rho }}{\partial{p}} 
\Theta \ , 
\mbox{\ \ \ \ \ \ \ \ \ \ \ \ \ \ \ }
\left(\tau = 0 \right)\ ,
\label{22}
\end{equation}
where we have combined Eqs. (\ref{6}) and (\ref{19}). 
Use of (\ref{18}) and (\ref{19}) in  (\ref{2}) yields 
\begin{equation}
\dot{\rho } = - \frac{\Theta }{3}\left(\rho + p \right)
\frac{\partial{\rho }}{\partial{p}} \ , 
\mbox{\ \ \ \ \ \ \ \ \ \ \ \ \ \ \ }
\left(\tau = 0 \right)\ .
\label{23}
\end{equation}
For a gas one has \cite{ZTP}
\begin{equation}
\frac{\partial{\rho }}{\partial{p}} = \left(\frac{m}{T} \right)^{2} 
- 1 + 5 \frac{\rho + p}{n T} 
- \left(\frac{\rho + p}{n T} \right)^{2}\ .
\label{24}
\end{equation}
Explicit integration of Eqs. (\ref{22}) and (\ref{23}) is possible in the 
limiting cases of radiation and nonrelativistic 
matter. 
Introducing a length scale $a$ according to 
\begin{equation}
\Theta \equiv  3 \frac{\dot{a}}{a}\ ,
\label{25}
\end{equation}
one recovers the dependences 
\begin{equation}
n \propto a ^{-3}\ , 
\mbox{\ \ \ } 
\rho \propto a ^{-4} \ , 
\mbox{\ \ \ \ }
\left(p = \rho /3 \ , \ \tau = 0 \right) 
\label{26}
\end{equation}
for radiation since $\Gamma $ as well as $\pi $ vanish in this case. 
For nonrelativistic matter on the other hand, one obtains 
\cite{ZGeq,ZiBa}
\begin{equation}
n \propto a ^{-3/2}\ , 
\mbox{\ \ \ }
\rho \propto a ^{-3/2} \ , 
\mbox{\ \ \ } 
\left(m \gg T \ ,\ \ 
\tau = 0 \right)\ .
\label{27}
\end{equation}
The temperature behaves according to (\ref{14}) 
for any equation of state. 
Since nonrelativistic matter is produced at a rate 
$\Gamma = \Theta /2$ [cf (\ref{21})] there are considerable deviations from the standard behavior for conserved particle number ( $n \propto a ^{-3}$, 
$T \propto a ^{-2}$, $\rho \propto a ^{-3}$).  
All thermodynamic quantities decrease more slowly than for $\Gamma = 0$ 
since the decay of 
$n$, $T$ and $\rho $ due to the expansion is counteracted by corresponding production terms. 
The differences of the behaviour of $n$, $T$ and $\rho$ acoording to the relations (\ref{27}) compared with the standard behaviour 
are 
consequences of the back reaction of the production process on the fluid dynamics. 
Also the expansion of the universe is modified. 
In a homogeneous and isotropic universe the length scale $a$ coincides with 
the scale factor of the Robertson-Walker metric. 
Restricting ourselves to the spatially flat case, the scale factor obeys 
the Friedmann equation 
\begin{equation}
3 \frac{\dot{a}^{2}}{a ^{2}} = \kappa \rho \ ,
\label{28}
\end{equation}
where $\kappa$ is Einstein's gravitational constant. 
For radiation we recover, of course, $a \propto t ^{1/2}$. 
But inserting the energy density $\rho $ of   
relations (\ref{27}) into (\ref{28}) we find that 
the scale factor behaves such as \cite{ZGeq,ZiBa}
\begin{equation}
a \propto t ^{4/3} \ , 
\mbox{\ \ \ \ \ \ \ \ \ \ \ \ \ \ \ }
\left(m \gg T \ ,\ \ \tau = 0 \right)\ ,
\label{29}
\end{equation}
i.e., $\ddot{a} > 0$ instead of the familiar $a \propto t ^{2/3}$ with  
$\ddot{a} < 0$ for $\rho \propto a ^{-3}$, corresponding to $\Gamma = 0$. 
The production of massive particles in equilibrium implies accelerated expansion ($\ddot{a}>0$), i.e., power-law inflation. 
In other words, generalized  equilibrium for nonrelativistic particles 
is only possible in a power-law inflationary universe.  

\section{The second-order theory}
\subsection{General relations}
In order to prepare the corresponding investigation of  
``generalized equilibrium ''  within the second-order theory we first realize  [cf (\ref{11})] that 
for $\Gamma = 0$ the first equilibrium condition in  (\ref{12}) is equivalent to 
\begin{equation}
\Theta \pi   + \frac{\tau }{\zeta }\pi \dot{\pi } 
+ \frac{\pi ^{2}}{2}\frac{\tau }{\zeta }
\left(\Theta + \frac{\dot{\tau }}{\tau } 
- \frac{\dot{\zeta }}{\zeta } - \frac{\dot{T}}{T} \right)
 = 0 \ .
\label{30}
\end{equation}
This condition may trivially be fulfilled for $\pi = 0$. 
However, there exists a different possibility, namely 
a compensation of the terms on the left-hand side of the last equation, 
\begin{equation}
\tau \dot{\pi } = - \zeta  \Theta - \frac{1}{2}\pi \tau 
\left[\Theta + \frac{\dot{\tau }}{\tau } 
- \frac{\dot{\zeta }}{\zeta } - \frac{\dot{T}}{T} \right]\ .
\label{31}
\end{equation}
We want to point out that this equation for $\pi $ does not coincide with the conventional, causal evolution equation for the corresponding  quantity of the Israel-Stewart theory (subscript IS) (see,e.g., \cite{WZ,Roy} and references therein), which we write down for comparison:
\begin{equation}
\pi _{_{\left(IS \right)}}  + \tau \dot{\pi }_{_{\left(IS \right)}} = - \zeta  \Theta 
- \frac{1}{2}\pi _{_{\left(IS \right)}}\tau 
\left[\Theta + \frac{\dot{\tau }}{\tau } 
- \frac{\dot{\zeta }}{\zeta } - \frac{\dot{T}}{T} \right]\ .
\label{32}
\end{equation}
The equations (\ref{31}) and (\ref{32}) refer to different physical situations. 
While Eq. (\ref{32}) corresponds to an entropy production 
\begin{equation}
S ^{a}_{;a} = \frac{\pi ^{2}_{_{\left(IS \right)}}}{\zeta T}\ ,
\label{33}
\end{equation}
which vanishes only for vanishing $\pi $, Eq. (\ref{31}) implies 
$S ^{a}_{;a} = 0$ although $\pi $ is different from zero. 
The causal evolution equation (\ref{32}) follows from the requirement to guarantee $S ^{a}_{;a} \geq 0$ in the simplest possible way 
with an entropy production given by the 
expression (\ref{33}).   
The viscous pressure $\pi _{_{\left(IS \right)}}$ in this case represents deviations from the equilibrium $S ^{a}_{;a} = 0$. 
In the present paper we are not interested in such kind of deviations from ``conventional'' equilibrium. 
Our point is to realize that ``generalized'' 
equilibrium within the second-order theory does not necessarily imply $\pi = 0$. 
There exists an alternative way to make the entropy production vanish, namely a nonvanishing $\pi $ satisfying Eq. (\ref{31}). 
This option relies on a compensation between the contributions to 
$S ^{a}_{;a}$ following from each of the two terms on the right-hand side of (\ref{7}). 
One cannot expect, of course, that this configuration implies an equilibrium distribution on the level of kinetic theory.   

The bulk pressure $\pi _{_{\left(IS \right)}}$ is a dynamical degree of freedom in (\ref{32}), the latter being  the corresponding evolution equation. 
Eq. (\ref{31}) on the other hand, which also may be written as 
\begin{equation}
\left(\frac{\tau }{\zeta } \right)^{^{\displaystyle \cdot}} 
+ \left(2 \frac{\dot{\pi }}{\pi } + \Theta - \frac{\dot{T}}{T}\right)
\frac{\tau }{\zeta } + 2\frac{\Theta }{\pi } = 0 \ ,
\mbox{\ \ \ }
\left(\pi \neq 0 \right)\ ,
\label{34}
\end{equation}
is an equation for the function $\tau /\zeta $. 
It will be shown below that 
the viscous pressure $\pi $ is  fixed by the CKV property of $u _{i}/T$, not yet taken into account in (\ref{31}), also in the second-order theory. 

Using the generalized equilibrium conditions (\ref{12}) in 
(\ref{11}) yields  
\begin{equation}
\Theta \pi   + \left(\rho + p \right)\Gamma 
+ \left(\frac{\pi ^{2}}{2}
\frac{\tau }{\zeta } \right)^{\displaystyle \cdot} 
+ \frac{2}{3} \pi ^{2}\frac{\tau }{\zeta }\Theta = 0 \ .
\label{35}
\end{equation}
Combination with Eqs. (\ref{10}) and (\ref{25})
allows us to write this condition as 
\begin{equation}
nT a ^{4}\dot{s} = \left(a ^{4}\frac{\pi ^{2}}{2}
\frac{\tau }{\zeta } \right)^{\displaystyle \cdot}\ .
\label{36}
\end{equation}
Let us now define a quantity 
\begin{equation}
s ^{*} = s - \frac{1}{2} \frac{\pi ^{2}}{nT}\frac{\tau }{\zeta }\ ,
\label{37}
\end{equation}
with the help of which we may write the entropy flow vector (\ref{7}) as 
$S ^{a} = n s ^{*}u ^{a}$. 
For $\tau \rightarrow 0$ the quantities $s ^{*}$ and $s$ coincide. 
Within the framework of Extended Irreversible Thermodynamics (EIT) 
\cite{JouCVLe,JerGer,Roy} a corresponding expression, 
\[
s ^{*}_{_{\left(IS \right)}} = s - \frac{1}{2} 
\frac{\pi ^{2}_{_{\left(IS \right)}}}{nT}\frac{\tau }{\zeta }\ ,
\]
is interpreted as nonequilibrium entropy per particle. 
Since $\pi $, different from $\pi _{_{\left(IS \right)}}$, does 
{\it not} describe deviations from (generalized) equilibrium (see the discussion following Eq. (\ref{33})), the quantity (\ref{37}) is a (generalized) equilibrium variable as well. 
It may be called ``second-order generalized equilibrium'' entropy per particle.  
 
With Eq. (\ref{37}) and $Ta = const$, following from 
(\ref{14}) and (\ref{25}), as well as  
$\Gamma = \dot{N}/N$, where $N = n a ^{3}$ is the number of particles in a  comoving volume $a ^{3}$, Eq. (\ref{36}) 
may be written as 
\begin{equation}
nT \dot{s}^{*} = \frac{\pi ^{2}}{2}\frac{\tau }{\zeta }\Gamma \ .
\label{38}
\end{equation}
It follows that in the limiting case $\Gamma = 0$ 
the quantity $s ^{*}$ remains constant along the fluid flow lines. As we will see below, this does not necessarily imply 
a vanishing $\pi $.

Comparing the rate of change of the temperature (\ref{16}) using 
$\dot{s}$ from (\ref{10}) with the corresponding rate 
(\ref{14}) 
we obtain 
\begin{equation}
\pi = \frac{1}{3}T \frac{\partial{\rho }}{\partial{T}}
\left(1 - 3 \frac{\partial{p}}{\partial{\rho }} \right) 
- n \frac{\partial{\rho }}{\partial{n}}\frac{\Gamma }{\Theta }\ ,
\label{39}
\end{equation} 
which replaces the expression (\ref{18}) of the first-order theory. 
Again, the generalized equilibrium conditions establish a relation between the rate $\Gamma $ and the bulk pressure $\pi $. 
While in the first-order theory $\pi $ vanishes with vanishing $\Gamma $, it is obvious from Eq. (\ref{39}) that $\pi $ will be nonvanishing in general even for 
$\Gamma = 0$. 
Only that part of $\pi $ which is proportional to $\Gamma $  is a 
``creation'' pressure. 
The first term on the right-hand side of (\ref{39}) is independent of the particle production and has to be regarded as a ``conventional'' viscous pressure. 
This structure of $\pi $ is a manifestation of the well-known fact that bulk viscosity and particle creation are independent phenomena which, however, may occur simultaneously \cite{GaLeDe}. 

\subsection{Particle conservation ($\Gamma = 0$)}
For vanishing  $\Gamma $ the effective viscous pressure (\ref{39}) reduces to 
\begin{equation}
\pi = \frac{1}{3}T \frac{\partial{\rho }}{\partial{T}}
\left(1 - 3 \frac{\partial{p}}{\partial{\rho }} \right) \ .
\label{40}
\end{equation}
This quantity vanishes for relativistic matter  
($p = \rho /3$). 
For any other equation of state, however, the equilibrium conditions require a nonvanishing $\pi $. 
While Eq. (\ref{35}) may also be satisfied with $\pi = 0$ for $\Gamma = 0$, 
Eq. (\ref{14}) may not. 
Consequently, there exists an equilibrium state (vanishing entropy production) within the second-order theory which admits a nonvanishing bulk pressure 
$\pi $. 
This apparently new result relies on the compensation effects mentioned below (\ref{30}) and in between (\ref{33}) and (\ref{34}). 
We recall that this case will certainly not be realised microscopically by a distribution function for collisional equilibrium. 
The part of $\pi $ which is not due to $\Gamma $  represents a 
``conventional'' bulk viscous pressure. 
These considerations show that within the second-order theory the set of equilibrium relations (\ref{12}) admits qualitatively new configurations compared with the first-order theory. 
In fact, the case $\Gamma =0$ discussed here is a consequence of the standard equilibrium requirement $S ^{a}_{;a}=0$  under the condition that 
$u _{a}/T$ be a conformal Killing vector. 

For nonrelativistic matter ($p = nT$, 
$\rho = nm + \frac{3}{2}nT$, $m \gg T$) one obtains 
\begin{equation}
\pi = - \frac{1}{2}nT \ ,
\label{41}
\end{equation}
i.e. $|\pi| = p/2$, although $S ^{a}_{;a} = 0$. 
Using the expression (\ref{41}) for $\pi $ in  (\ref{37}) the property $\dot{s}^{*} = 0$ 
[cf (\ref{38})] is equivalent to 
\begin{equation}
\left(nT \frac{\tau }{\zeta } \right)^{\displaystyle \cdot} 
= 12 \frac{\dot{a}}{a}\ .
\label{42}
\end{equation}
Integration yields 
\begin{equation}
\frac{\tau }{\zeta } = \frac{12}{nT} 
\ln \left(\frac{a}{a _{0}} \right)\ .
\label{43}
\end{equation}
The quantity $\tau / \zeta $ is only determined up to a constant. 
We have used here this freedom to make $\tau / \zeta $ vanish for some initial value 
$a = a _{0}$. 
Since $\tau / \zeta $ vanishes identically in the first-order theory this may be interpreted as ``switching on'' the second order at $a = a _{0}$. 
The function $\tau / \zeta $ then describes deviations from the corresponding first-order description. 
In the Israel-Stewart theory the function $\tau / \zeta $ is known to be related to the viscous part 
$c _{b} $ of the sound velocity in a viscous fluid by \cite{HL,RM}
\begin{equation}
c _{b}^{2} = \frac{1}{\rho + p}\frac{\zeta }{\tau }\ .
\label{44}
\end{equation}
This expression follows also in the appropriate limit from formula (59) of \cite{GaCi} with the identifications $nF \rightarrow \rho + p$ and 
$\beta _{0} = \frac{\tau }{n \zeta }$. 
It is remarkable that the system of perturbation equations (53) - (56) in 
\cite{GaCi} from which the velocity (\ref{44}) is obtained, does not change if Eq. (\ref{32}) of the general analysis is replaced by Eq. (\ref{31}). 
Therefore, the expression (\ref{44}) may be used in the present context as well. 
We associate $c _{b} $ with the effective viscous part of a propagation velocity in a fluid in generalized equilibrium. 
The quantity $c _{b}^{2}$ diverges for $\tau \rightarrow 0$ (vanishing second order) reflecting the ``acausal'' nature of the first-order theories. 
The more familiar adiabatic part of the sound velocity $c _{s}$ is given by
\begin{equation}
c _{s}^{2}  = \left(\frac{\partial{p}}{\partial{\rho }} \right)_{ad} 
= \frac{n}{\rho + p}\frac{\partial{p}}{\partial{n}} 
+ \frac{T}{\rho + p} 
\frac{\left(\partial p / \partial T \right)^{2}}
{\partial \rho / \partial T}\ .
\label{45}
\end{equation}
The sound in a viscous medium propagates with a subluminal velocity $v$ 
where \cite{RM}
\begin{equation}
v ^{2} = c _{s}^{2} + c _{b}^{2}  \leq 1\ .    
\label{46}
\end{equation}
With the equations of state for nonrelativistic matter we find 
\begin{equation}
c _{s}^{-2} = \frac{3}{5}\frac{m}{T}\ ,
\mbox{\ \ \ \ \ \ \ }
c _{b}^{-2}  = 12 \frac{m}{T}\ln \left(\frac{a}{a _{0}} \right)\ .
\label{47}
\end{equation}
For $a$ sufficiently larger than $a _{0}$ (recall that $a = a _{0}$ represents the (unphysical) first-order limiting case) we have 
$c _{s}^{2} \ll 1 $, $c _{b}^{2} \ll 1 $ and 
$c _{b}^{2} < c _{s}^{2} $, i.e., the ``adiabatic'' part exceeds the viscous one which is consistent with $| \pi | < p$ according to 
(\ref{41}). 
This completes our discussion of equilibrium states 
of the second-order theory for the case of vanishing particle production. 

\subsection{Particle production ($\Gamma \geq 0$)}
According to (\ref{38}) the generalized entropy per particle $s ^{*}$ 
does not remain constant for $\Gamma > 0$. 
Taking into account Eq. (\ref{37}) and $N = n a ^{3}$, Eq. (\ref{38}) may be written as 
\begin{equation}
\left[N \left(s ^{*} - s \right) \right]^{\displaystyle \cdot} 
= - N \dot{s}\ .
\label{48}
\end{equation}
The first-order case corresponds to $s ^{*} = s$, implying $\dot{s} = 0$. 
We will integrate Eq. (\ref{48}) for the limiting cases of pure radiation and nonrelativistic matter. 
In the first case ($p = n T$, $\rho = 3 n T$) the expression (\ref{39}) for the viscous pressure reduces to 
\begin{equation}
\pi = 
- n \frac{\partial{\rho }}{\partial{n}}\frac{\Gamma }{\Theta } 
= - 3nT \frac{\Gamma }{\Theta }\ .
\label{49}
\end{equation} 
Use of the last equation together with relation (\ref{17}) and $p = n T$ in 
(\ref{10}) yields
\begin{equation}
N \dot{s} = - N \Gamma = - \dot{N}
\label{50}
\end{equation}
for the right-hand side of (\ref{48}) which, 
consequently, may be written as
\begin{equation}
\left[N \left(1 + \frac{1}{2}\frac{\pi ^{2}}{nT}
\frac{\tau }{\zeta } \right) \right]^{\displaystyle \cdot} = 0 \ ,
\label{51}
\end{equation}
with $\pi $ from (\ref49{}). 
Assuming again the funtion $\tau / \zeta $ to vanish for some initial time, 
i.e. $\left(\tau / \zeta  \right)_{0} = 0$, the solution of (\ref{51}) 
is 
\begin{equation}
- \frac{1}{2}\frac{\pi ^{2}}{nT}\frac{\tau }{\zeta } 
= \frac{N - N _{0}}{N}\ ,
\label{52}
\end{equation}
where $N _{0}$ is the initial particle number. 
For $N > N_{0}$ the last relation may only be fulfilled if the ratio 
$\tau / \zeta $ is negative. 
Since formula (\ref{44}) necessarily implies a nonnegative value of this ratio,  we conclude that the only physical solution of 
(\ref{51}) is $\pi = \Gamma = 0$. 
This confirms the corresponding result of the first-order theory according to which the production of relativistic particles is incompatible with the conditions of generalized equilibrium. 

For nonrelativistic matter 
($p = nT$, $\rho = n m + \frac{3}{2}nT$, $m \gg T$) the viscous pressure (\ref{39}) becomes 
\begin{equation}
\pi = - \frac{1}{2} n T - \rho \frac{\Gamma }{\Theta }\ ,
\mbox{\ \ \ \ \ \ \ \ \ \ \ \ }
\left(m \gg T \right)\ .
\label{53}
\end{equation}
For $\Gamma = 0$ we recover the limiting case (\ref{41}). 
Introducing the expression (\ref{53}) into (\ref{10}) 
yields 
\begin{equation}
\dot{s} = \frac{1}{2}\Theta - \Gamma \ , 
\mbox{\ \ \ \ \ \ \ \ \ \ \ \ }
\left(m \gg T \right)
\ .
\label{54}
\end{equation}
The condition $\dot{s} = 0$ correctly reproduces the relation 
$\Gamma = \Theta /2$ [cf (\ref{21})] of the  first-order theory. 
For $N \dot{s}$ on the right-hand side of 
(\ref{48}) we obtain 
\begin{equation}
N \dot{s} = \frac{3}{2} N \frac{\dot{a}}{a} - \dot{N} \ , 
\mbox{\ \ \ \ \ \ \ \ \ \ \ } 
\left(m \gg T \right)
\ .
\label{55}
\end{equation}
To integrate this expression we assume $\Gamma $ to have the form 
\begin{equation}
\Gamma = \frac{\beta }{2}\Theta \ , 
\mbox{\ \ \ \ \ \ \ \ \ \ \ \ } 
\left(m \gg T \right)\ ,
\label{56}
\end{equation}
with a constant parameter $\beta $. The case $\beta = 1$ corresponds to the first-order theory. 
Differences between the second- and first-order theories are then characterized by deviations from $\beta = 1$. 
With the ansatz (\ref{56}) integration of the particle number balance 
(\ref{6}) yields
\begin{equation}
N = N _{0}\left(\frac{a}{a _{0}} \right)^{3 \beta /2} 
\mbox{\ \ \ \ \ \ \ \ \ \ \ \ } 
\left(m \gg T \right)
\ .
\label{57}
\end{equation}
Using this law for $N$ in Eq. (\ref{55}), integration of the latter results in 
\begin{equation}
\int_{t _{0}}^{t} N \dot{s}\mbox{d}t = \frac{1 - \beta }{\beta }
\left(N - N _{0} \right)\ ,
\ \ \ \ 
\left(m \gg T \right)
\ .
\label{58}
\end{equation}
Applying now the definition 
(\ref{37}) of $s ^{*}$ and assuming again 
$\tau / \zeta $ to vanish initially 
(see the discussion following (\ref{43})), the integral of 
(\ref{48}) 
turns out to be 
\begin{equation}
\frac{\pi ^{2}}{2nT}\frac{\tau }{\zeta } = \frac{1 - \beta }{\beta }
\frac{N - N _{0}}{N}\ .
\label{59}
\end{equation} 
The limit $\beta = 1$, i.e. $\Gamma = \Theta /2$, corresponds to 
$\tau = 0$ of the first-order theory. 
The case $\beta > 1$ is unphysical since it is incompatible with a positive 
$\tau / \zeta $. 
Consequently, the second-order theory requires $\beta < 1$. 
In other words, the generalized equilibrium conditions of the second-order theory imply a production rate which is less than the corresponding rate of the first-order theory. 
To be more specific, we assume $\beta $ to be smaller but still of the order of 1. 
In such a case $\pi $ is approximately given by 
\begin{equation}
\pi \approx - \frac{\beta }{2}n m \ ,
\mbox{ \ \ \ \ \ \ \ \ \ \ \ \ }
\left(m \gg T \right)
\ , 
\label{60}
\end{equation}
i.e., it is entirely determined by the particle production. 
Use of $\tau / \zeta  = 
\left[\left(\rho + p \right)c _{b}^{2} \right]^{-1} 
\approx \left[n m c _{b}^{2} \right]^{-1}$ [cf (\ref{44})] and 
relation (\ref{60}) allows us to write the left-hand side of 
(\ref{59}) as 
\begin{equation}
\frac{1}{2}\frac{\pi ^{2}}{nT}\frac{\tau }{\zeta } 
= \frac{\beta ^{2}}{8 c _{b}^{2}}\frac{m}{T}\ ,
\mbox{ \ \ \ \ \ \ \ \ \ \ \ \ }
\left(m \gg T \right)
\ .
\label{61}
\end{equation}
Again, the case $\tau \rightarrow 0$ corresponds to the (unphysical) 
behaviour $c _{b}^{2} \rightarrow \infty$. 
Combining Eqs. (\ref{59}) and (\ref{61}) and assuming $N \gg N _{0}$, i.e. considerable particle production, we obtain
\begin{equation}
c _{b}^{2} = \frac{1}{8}\frac{\beta ^{3}}{1 - \beta }\frac{m}{T}\ ,
\mbox{ \ \ \ \ \ \ \ \ \ \ \ \ }
\left(m \gg T \right)
\ .
\label{62}
\end{equation}
Recalling from Eq. (\ref{47}) 
that the adiabatic part of the sound velocity is characterized  by  
$c _{s}^{2} \ll 1$ for $T \ll m$, the condition $v ^{2} \leq 1$,  
restricting the sound propagation to subluminal speeds, reduces to 
\begin{equation}
c _{b}^{2} < 1 \quad\Rightarrow\quad 
\frac{m}{T} < \frac{8 \left(1 - \beta  \right)}{\beta ^{3}}\ .
\label{63}
\end{equation}
The latter inequality has to be fulfilled for 
$m/T \gg 1$. 
These simultaneous requirements restrict the possible $\beta $-values and, according to (\ref{56}) limit the production rate. 
While the generalized equilibrium conditions of the first-order theory 
imply the result $\beta = 1$ for nonrelativistic particles, $\beta $ is not  uniquely fixed within the second-order theory. 
The conditions $m \gg T$ and (\ref{63}) determine only a range of possible 
$\beta $-values. 
With $\beta = 1/2$, e.g., we find 
$8 \left(1 - \beta  \right)/\beta ^{3} = 32$, a value perfectly admissible 
for $m/T$. 
For $\beta = 2/3$ one has $m/T < 9$ which means that $m$ still may be almost an order of magnitude larger than $T$. 
For $\beta = 3/4$ we obtain $m/T < 4.74$, a value already hardly compatible with 
$m \gg T$. 
For values of $\beta $ still closer to 1 the upper limit for $m/T$ becomes even lower. 
We conclude that the causality requirement of the second-order theory restricts $\beta $ to values of the order of $2/3$, i.e. 
\begin{equation}
\beta \leq \frac{2}{3 } 
\quad\Rightarrow\quad \Gamma \leq \frac{\Theta }{3} 
\quad\Rightarrow\quad |\pi| \leq \frac{1}{3}nm \ ,
\ \ \ \ 
\left(m \gg T \right)
\ .
\label{64}
\end{equation}
The limiting value $\Gamma  = \Theta /3$ corresponds to 
$\pi = - nm/3$. 
Use of the latter in the energy balance (\ref{2}) implies 
\begin{equation}
\rho \propto a ^{-2}\ ,
\mbox{ \ \ \ \ \ \ \ \ \ \ \ \ }
\left(m \gg T \right)
\ .
\label{65}
\end{equation}
The second-order theory predicts $\rho $ to decay faster than the 
first-order theory [cf (\ref{27})]. 
A behaviour of the energy density such as given by Eq. (\ref{65}) is characteristic for so-called ``K-matter'' \cite{Kolb}. 
Consequently, generalized equilibrium for massive particles within the second-order theory may be regarded as realization of K-matter. 
In a spatially flat universe the behaviour (\ref{65}) implies 
$a \propto t$, i.e. $\dot{a} = 0$ (coasting cosmology). 
While for massive particles the generalized equilibrium of the first-order theory is connected with power-law inflation $a \propto a ^{4/3}$, i.e. accelerated expansion, the second-order theory reduces the acceleration of the expansion.   
The first-order theory uniquely fixes the power $q$ in $a \propto t ^{q}$ for 
the scale factor $a$ of a spatially flat universe of massive particles to 
$q = 4/3$. 
The second-order theory restricts this power to $q \leq 1$. 
In other words, the first-order theory overestimates the generalized equilibrium value for the particle production and the back reaction of the latter on the cosmological dynamics. 

\section{Summary}
We have defined generalized equilibrium in second-order thermodynamics by the set of conditions (\ref{12}) together with the expression (\ref{7}). 
For $p = \rho /3$ (radiation) this kind of equilibrium reduces to the standard 
``global '' equilibrium of relativistic thermodynamics and requires particle number conservation both in the first- and second-order theories. 
For any other equation of state the first- and second-order generalized equilibrium states are different. 
While the first-order equilibrium necessarily implies nonzero particle production (except for $p = \rho /3$) the second-order theory admits (generalized) equilibrium both for vanishing and nonvanishing production rates. 
In the first case (no particle production) vanishing entropy production is compatible with a nonvanishing viscous pressure. 
In the second case (nonzero particle production) the equilibrium particle production rate is lower than the corresponding first-order quantity. 
Consequently, the back reaction of the production process on the cosmological dynamics is weaker than predicted on the basis of the first-order theory. 
Especially, the property of power law inflation $a \propto t ^{4/3}$ in a spatially 
homogeneous and flat universe filled with nonrelativistic matter in first-order generalized equilibrium does no longer hold in the second-order theory.  
The maximally possible production rate in the second-order theory is of the order of the Hubble expansion rate $H \equiv \Theta /3$ compared with $\Theta /2$ of the first-order theory. 
This limiting value ($\Gamma \approx H$ for nonrelativistic matter) 
of the second-order theory implies the dependence $\rho \propto a ^{-2}$ 
($\dot{a}$ = const) for the energy density, a behavior characteristic for 
K-matter. 
A universe of massive particles in generalized second-order equilibrium may realize a ``coasting'' cosmology. \\
\ \\
{\bf Acknowledgements}\\
This paper was supported by the CNRS and the Deutsche Forschungsgemeinschaft. 
Discussions with Diego Pav\'{o}n are gratefully acknowleged. 
W.Z. thanks the Laboratoire de Gravitation et Cosmologie Relativistes, Universit\'{e} Pierre et Marie Curie, Paris for warm hospitality.
\ \\

\end{document}